\documentstyle[emulateapj,graphicx]{article}
\def\gsim{\;\rlap{\lower 2.5pt
 \hbox{$\sim$}}\raise 1.5pt\hbox{$>$}\;}
\def\lsim{\;\rlap{\lower 2.5pt
   \hbox{$\sim$}}\raise 1.5pt\hbox{$<$}\;}
\newcommand{\redshift} {\left( \frac{1+z}{10} \right)}
\newcommand{\redshiftt} {[(1+z)/10]}
\newcommand{\SFR} {{\rm \left( \frac{SFR}{1 M_{\odot} \, yr^{-1}} \right)}}  
\newcommand{\SFRt} {{\rm (SFR/M_{\odot}\, yr^{-1})}}

\newcommand{\beq}{\begin{equation}}
\newcommand{\eeq}{\end{equation}}
\def\myputfigure#1#2#3#4#5%
{\hskip0.03\textwidth\vskip#5pt
\makebox[0pt]{\hskip#2in
\includegraphics[width=#3\textwidth]{#1}}\vskip#4pt\hfill}


\lefthead{Oh, Haiman \& Rees}
\righthead{HeII Recombination Lines From the First Objects}

\begin{document}
\title{HeII Recombination Lines From the First Luminous Objects}
\author{S. Peng Oh, Zolt\'an Haiman\altaffilmark{1}}
\affil{Princeton University Observatory, Princeton, NJ 08544, USA\\
peng@astro.princeton.edu, zoltan@astro.princeton.edu}
\vspace{0.5\baselineskip}
\author{Martin J. Rees}
\vspace{0.1\baselineskip}
\affil{Institute of Astronomy, Madingley Road, Cambridge, CB3 0HA, UK\\
mjr@ast.cam.ac.uk}
\altaffiltext{1}{Hubble Fellow}

\begin{abstract}
The hardness of the ionizing continuum from the first sources of UV radiation
plays a crucial role in the reionization of the intergalactic medium
(IGM). While usual stellar populations have soft spectra, mini--quasars or
metal--free stars with high effective temperatures may emit hard photons,
capable of doubly ionizing helium and increasing the IGM temperature.
Absorption within the source and in the intervening IGM will render the
ionizing continuum of high--redshift sources inaccessible to direct
observation.  Here we show that HeII recombination lines from the first
luminous objects are potentially detectable by the {\it Next Generation Space
Telescope}. Together with measurements of the H$\alpha$ emission line, this
detection can be used to infer the ratio of HeII to HI ionizing photons,
$Q=\dot{N}_{\rm ion}^{\rm HeII}/\dot{N}_{\rm ion}^{\rm HI}$.  A measurement of
this ratio would shed light on the nature and emission mechanism of the first
luminous sources, with important astrophysical consequences for the reheating
and reionization of the IGM.
\end{abstract}

\section{Introduction}

The launch of the {\it Next Generation Space Telescope (NGST)} raises the
exciting possibility of direct imaging and spectroscopy of the first luminous
objects that reionized the universe at high redshifts $z\sim 5-30$. The first
mini--galaxies are likely detectable in rest frame UV emission (Haiman \& Loeb
1997), assuming a cold dark matter (CDM) cosmology, and an average star
formation efficiency calibrated to the observed metallicity of the Ly$\alpha$
forest at redshift $z\sim 3$.  Similarly, mini--quasars with black hole (BH)
masses comparable to those found in nearby galaxies (Gebhardt et al. 2000;
Ferrarese \& Merritt 2000) and shining at the Eddington limit would be
observable (Haiman \& Loeb 1998). Detection of these sources before the epoch
of reionization in the Ly$\alpha$ emission line is difficult due to the high
optical depth of the intervening neutral IGM. This causes the Ly$\alpha$
photons to be spread out over an extended low surface brightness halo (Loeb \&
Rybicki 1999), although if the first ionizing sources were very luminous
quasars, they may ionize a sufficiently large surrounding region that
Ly$\alpha$ detection becomes feasible (Cen \& Haiman 2000, Madau \& Rees
2000). By comparison, resonant scattering is unimportant for Balmer line
emission, which is likely to be detectable by {\it NGST} (Oh 1999), even for
sources which do not ionize their surroundings out to great distances.

The above observations do not strongly constrain the hardness of the ionizing
spectrum. Hard sources of ionizing radiation are certainly plausible at high
redshift, and can vary from mini--quasars (Haiman \& Loeb 1998), to very
massive objects ($\gsim 100~{\rm M_\odot}$ VMOs, Bond, Carr, \& Arnett 1984),
metal--free stars (Castellani et al. 1983; El Eid et al. 1983; Tumlinson \&
Shull 2000), thermal and non-thermal emission from supernova remnants (Oh
2000a), X-ray binaries, and so forth. In particular, while low-metallicity ($Z
\sim 10^{-2} {\rm Z_{\odot}}$) stellar populations with a Salpeter or Scalo IMF
(which are often used in reionization models) emit a negligible amount of HeII
ionizing photons, this is not the case for these other sources of ionizing
radiation.

The spectral hardness of the first sources of ionizing radiation has important
consequences for cosmological reionization.  If the first ionizing sources emit
copious numbers of HeII ionizing photons, HeII might be substantially reionized
at high redshifts.  The fast recombination time of HeII allows efficient
transfer of energy from the radiation field to the IGM, raising the IGM to
significantly higher temperatures (Miralda-Escud\'e \& Rees 1994).  In
addition, if the ionizing spectrum extends to X-ray energies ($\sim 1$keV), the
hard photons promote ${\rm H_2}$ molecule formation and cooling inside
minihalos, by increasing their free electron fraction (Haiman, Abel \& Rees
2000).  As a result, reionization could commence at higher redshifts. Finally,
due to their large mean free path, hard X--ray photons also change the topology
of reionization, making it gradual and homogeneous (Oh 2000b).

Despite its importance, the ionizing continuum of high--redshift sources will
likely be inaccessible to direct observation, because of strong absorption by
neutral H and He within the source, and also in the neutral IGM in which the
sources are embedded. In particular, the very high optical depth to resonant
scattering in the IGM by neutral hydrogen implies that flux shortward of
rest-frame Ly$\alpha$ is sharply attenuated (Gunn \& Peterson 1965). The
production rate of ionizing photons may be indirectly inferred by H$\alpha$
observations with NGST. As shown recently by Tumlinson \& Shull (2000), the
composite spectrum of a metal--free stellar population is significantly harder
than that of its low--metallicity counterpart: as a consequence, HeII
recombination lines from metal--free stars might also be observable with {\it
NGST}. The IGM is optically thin to these recombination lines, since the
de-excitation time for level transitions is very short and HeII always resides
in the ground state. In this paper, we show that that observing HeII
recombination lines from the first ionizing sources is feasible, and can serve
as a general diagnostic of any hard source of ionizing radiation. A measurement
of the strength of these emission lines can help distinguish between different
reionization scenarios: e.g. whether mini--galaxies or mini--quasars were the
dominant sources of ionizing radiation at high redshift.

\section{Detectability of HeII recombination lines}

The escape fraction $f^{\rm HI}_{\rm esc}$ of HI ionizing photons from galaxies
in the local universe is inferred to be small, $\sim 3-6 \%$ (Leitherer et al
1995, Bland-Hawthorn \& Maloney 1999, Dove, Shull \& Ferrara 2000) and is
expected to further decrease with increasing redshift (Wood \& Loeb 2000,
Ricotti \& Shull 2000).  The escape fraction $f^{\rm HeII}_{\rm esc}$ of HeII
ionizing photons is likely to be even smaller: all the sources of ionizing
radiation we consider have $Q \equiv \dot{N}^{\rm HeII}_{\rm ion}/\dot{N}^{\rm
HI}_{\rm ion} \ll (n_{\rm He}/n_{\rm H}) (t_{\rm rec,HI}/t_{\rm rec,HeII}) =
(\alpha_{\rm He II} n_{\rm He})/(\alpha_{\rm HI}n_{\rm H}) = 0.5$ (here
$\dot{N}^{\rm HI}_{\rm ion}$ and $\dot{N}^{\rm HeII}_{\rm ion}$ are the
ionizing photon emission rates above 13.6 and 54.4eV; $t_{\rm rec,HI}$ and
$t_{\rm rec,HeII}$ are the recombination timescales for HI and HeII, and
$\alpha_{\rm HI}$ and $\alpha_{\rm He II}$ are the case B recombination
coefficients). We thus expect that the extent of HeIII regions within the
galaxy is smaller than the extent of HII regions, implying $f^{\rm HeII}_{\rm
esc} < f^{\rm HI}_{\rm esc}$.

We proceed to estimate the luminosity of sources in recombination line emission
by assuming that the escape fraction $f_{\rm esc}$ of ionizing photons is
negligibly small.  In this case, under a wide range of nebulosity conditions,
the emission rate of recombination line photons is proportional to the emission
rate of ionizing photons (more generally, the line fluxes we predict would also
scale linearly with $\propto [1-f_{\rm esc}]$).  It is useful to express the
line intensities relative to the strongest line, the Balmer $\alpha$ line of
hydrogen.  For the Balmer $\alpha$ line, $\dot{N}_{\rm H\alpha} \approx 0.45
\dot{N}_{\rm ion}^{\rm HI}$ (Hummer \& Storey 1987). The luminosity in a line
$i$ is given by $L_{i}= 4 \pi j_{i} V = (4\pi j_{i})/(n_{e} n_{k})\times
(\dot{N}_{{\rm ion}}^{k})/(\alpha_{B,k})$, where $4\pi j_{i}$ is the emission
coefficient (in ${\rm erg~cm^{-3}~s^{-1}}$), $V$ is the emitting volume, and we
have assumed ionization balance for species $k$, which implies $\dot{N}_{{\rm
ion}}^{k}= \alpha_{B,k} n_{e} n_{k} V$. Thus, we can normalize the luminosity
of a helium recombination line $i$ to the H$\alpha$ luminosity via $L_{\rm i} =
Q f_{\rm i} L_{\rm H\alpha}$, where $L_{\rm H\alpha}$ is the H$\alpha$ line
luminosity, $f_{\rm i}\equiv \frac{j_{i}}{j_{H \alpha}} \left[ \frac{n_{\rm
HII}\alpha_{B,{\rm HI}} }{n_{\rm HeIII}\alpha_{B,{\rm HeII}} } \right]$, and we
obtain the quantities $j_{H \alpha}/n_{\rm HII}n_e$ and $j_{i}/n_{\rm
HeIII}n_e$ from Seaton (1978). Note that the similar temperature scaling of the HI, HeII emission and
recombination coefficients implies that $f_{\rm i} \propto T^{0.15}$ depends
only weakly on the assumed temperature, varying by at most $\sim 10 \%$ across
a plausible temperature range. We are only interested in the HeII recombination
lines longward of HI Ly$\alpha$, which will not redshift into HI Ly$\alpha$ and
be resonantly scattered in the IGM. The three strongest such recombination
lines are: HeII$(4 \rightarrow 3)$, HeII$(3 \rightarrow 2)$, and HeII$(5
\rightarrow 3)$, with rest frame wavelengths $\lambda = 4686 {\rm \AA}$, $1640
{\rm \AA}$, $3203 {\rm \AA}$, and luminosities relative to the H$\alpha$ line
of $f_{\lambda 4686} =0.74$, $f_{\lambda 1640}=4.7$, $f_{\lambda 3203} = 0.30$
respectively.

We now consider two examples for the detectability of the HeII recombination
lines: a metal--free stellar population with a Salpeter IMF, or a mini--quasar
with a BH of mass $M_{\rm bh}$.  For metal--free stars the ionizing photon
production rates are given by $\dot{N}_{\rm ion}^{\rm HI} \approx 1.5 \times
10^{53} \SFRt \, {\rm photons \,\, s^{-1}}$, and $N_{\rm ion}^{\rm HeII}\approx
10^{51.8} \SFRt \, {\rm photons \,\, s^{-1}}$, implying $Q\approx 0.05$ (
Tumlinson \& Shull, 2000). Thus, the flux from the He recombination line $i$ is
\begin{eqnarray}
\nonumber 
J_{\rm i}^{\rm stars} = && 
\frac{L_{\rm i}}{4 \pi d_{\rm L}^{2}} 
\frac{1}{\delta \nu} 
\approx 26 
\left( \frac{q_{\rm i}}{1.2} \right) 
\left( \frac{R}{1000} \right) 
\redshift^{-1} \times \\ 
&& 
\left( \frac{Q} {0.05} \right) 
\SFR \left( \frac{1-f_{esc}^{HeII}}{1} \right)
{\rm nJy},
\label{eq:Jstar}
\end{eqnarray}  
where $q_{\rm i} \equiv f_{\rm i} \nu_{\rm H\alpha}/ \nu_{\rm i}$ (so
that $q_{\lambda 4686}= 0.53, q_{\lambda 1640}=1.2, q_{\lambda 3203}=0.15$), $R \equiv
\lambda/ \delta \lambda$ is the spectral resolution, and SFR is the star
formation rate.  We have assumed that the line is unresolved, so that the
measured flux is linearly proportional to the spectral resolution.  The
intensity of the HeII lines is only a few percent of the flux from H$\alpha$
emission. For mini--quasars emitting ionizing radiation with a power law
spectrum $L_{\nu} \propto \nu^{-\alpha}$, we have $Q=4^{-\alpha}$.  The median
spectrum of Elvis et al. (1994) suggests $\alpha\approx 1$, while the
observations of Zheng et al. (1997) suggest $\alpha\approx 1.8$ (note that the
latter was observed only in the radio-quiet AGN subsample at energies up to 2.6
Ry, short of the 4.0 Ry HeII edge). We infer an implied range of $Q=0.08-0.25$,
so that a somewhat larger fraction of the ionizing radiation from quasars is
able to ionize HeII. Assuming the BH emits radiation at the Eddington rate, we
find the He recombination flux:
\begin{eqnarray}
\nonumber J_{\rm i}^{\rm quasar} 
\approx && 
56 
\left(\frac{q_{\rm i}}{1.2} \right) 
\left(\frac{R}{1000}\right) 
\redshift^{-1}
\left(\frac{M_{\rm bh}}{10^{5} \, M_{\odot}} \right)\times \\ 
&&
\left( \frac{4^{-\alpha}} {0.25} \right)  \left( \frac{1-f_{esc}^{HeII}}{1} \right)
{\rm nJy}.
\label{eq:Jquasar}
\end{eqnarray}   
Note that $f_{esc}^{HeII}$ for QSOs can be somewhat larger than that for
stars. Since the central black hole is a point source, it can create an ionized
chimney of gas perpendicular to the disk through which ionizing photons can
escape. This is also the case if gas is clumpy and the covering factor of
neutral clouds is low. However, unless the escape fractions are very high
($f_{esc}^{HeII} > 50 \%$), this does not significantly affect our estimates.
 
Despite the fact that HeII line flux is a small fraction of the H$\alpha$ line,
the HeII line can still be detected.  To estimate the signal--to--noise ratio
of the emission lines, we use the the expression for the noise expected for
{\it NGST} given by Gillett \& Mountain (1998, page 44, see also Oh 1999). In
the relevant range of fluxes and wavelengths (a few nJy at $1 \mu$m
$\lsim\lambda$ $\lsim 10\mu$m), the noise is expected to be dominated by the
detector dark current and the sky background. In the 1-5.5 $\mu$m range we
adopt a dark current of 0.02 $e$ ${\rm s^{-1}}$, based on the current design
goal of {\it NGST}.  The detector noise at $\lambda>5.5\mu$m is expected to
increase substantially; we assume a dark current of 0.3 $e$ ${\rm s^{-1}}$ in
this range based on the present state--of--the art detector on the {\it Space
Infrared Telescope Facility (SIRTF)}.  To estimate the sky background, we use
the minimum observed sky brightness tabulated from DIRBE data in the range
$1.25\mu$m-$12\mu$m (Hauser et al. 1998, Table 2).  We find that this
background contributes $\sim 20-50\%$ of the noise in the 1-5.5 $\mu$m range,
while it is less important ($\lsim 10\%$) compared to the increased detector
noise at $\lambda>5.5\mu$m. The sky background is dominated by the zodiacal
light, and might be substantially reduced for ${\it NGST}$ compared to DIRBE ,
depending on ${\it NGST}$'s orbit.

The increase of the detector noise beyond $\lambda>5.5\mu$m affects the
relative detectability of the HeII and H$\alpha$ lines. For a source at
redshift $z_{\rm s,i} < 32, 16, 11$, the He $\lambda 1640$, $\lambda 3203$,
$\lambda 4686$ lines lie in the favorable 1-5.5 $\mu$m range; by contrast, the
H$\alpha$ line redshifts to $\lambda > 5.5 \, \mu$m for $z_{\rm s} > 7$. Thus,
in the redshift range $7 < z < z_{\rm s,i}$ the relative signal to noise is
${\rm (S/N)_{He,i}/(S/N)_{H\alpha}} \approx 0.5 (Q/0.05) (q_{\rm i}/1.2)$. As
an example, for a metal--free stellar population with a Salpeter IMF, the
detection of the brightest line, HeII $\lambda 1640$, would require an
integration time $\approx$ 4 times longer than that required for H$\alpha$; a
quasar with $\alpha=1$ would an integration time $\sim 6$ times shorter for HeII
$\lambda 1640$ than for H$\alpha$.  More generally, for redshifts below $z_{\rm s,i}$, a
day's integration for a HeII line yields
\begin{eqnarray}
\nonumber
{\rm \frac{S}{N}} \approx 10 &&
\left(\frac{q_{\rm i}}{1.2}\right)
\left( \frac{R}{1000} \right) 
\redshift^{-1} 
\left(\frac{t}{10^{5} \, {\rm s}}\right)^{1/2} 
\times \\
&&
\nonumber
\left(\frac{\Omega}{0.01 \, {\rm arcsec^{-2}} } \right)^{-1/2}
\left[
\left(\frac{Q}{0.05}\right),
\left( \frac{4^{-\alpha}} {0.25} \right)\right]  
\times \\
&&
\left[
\left(\frac{\rm SFR}{1 {\rm M_{\odot}}\,{\rm yr^{-1}}}\right), 
\left(\frac{M_{\rm bh}}{5 \times 10^{4}\, {\rm M_{\odot}}}\right) \right] 
\label{eq:SN}
\end{eqnarray}
where the left figures in square brackets are to be used for metal--free stars
and the right figures are to be used for mini--quasars. Assuming that $f_{{\rm
star}} \sim 10\%$ of the gas is processed into stars, and $f_{{\rm BH}} \sim
0.3\%$ collapses to the central BH (see the models for high-redshift starbursts
and mini--quasars by Haiman \& Loeb 1997, 1998), the requirement of S/N=10
corresponds to dark halos with total mass $M_{\rm halo} \sim 10^{9} \, {\rm
M_{\odot}}$ at $z=9$ (note that $f_{{\rm star}},f_{{\rm BH}}$ should be
regarded as adjustable parameters uncertain to within an order of
magnitude). We assume that gas in sources cools and collapses to form discs
with angular scale $r_{\rm disc} \sim \lambda r_{\rm vir}$, where $\lambda \sim
0.05$ is the spin parameter; thus, most photons are converted to recombination
line photons in the dense cores of halos. In this case most sources remain
spatially unresolved, and subtend a solid angle $\Omega \approx {\rm max}[1,
(\lambda/3.5 \mu{\rm m})^{2}]$ 0.01 ${\rm arcsec}^{2}$.

From the observed recombination line fluxes, one can determine the relative
number of HI and HeII ionizing photons: $Q_{\rm obs} = (1/q_{\rm i}) (J_{\rm
HeII,i}/J_{\rm H\alpha})$.  There is, however, an upper limit to the observable
ratio of HeII and H$\alpha$ line strengths.  Ionizing photon ratios in excess
of $\dot{N}_{\rm ion}^{\rm HeII}/\dot{N}_{\rm ion}^{\rm HI} \sim 1$ do not
further increase the observed line ratios, because $\approx 50\%$ of the
opacity at $E=54.4$eV is contributed by HI rather than HeII, inhibiting the
growth of a HeIII region further than the extent of the HII region.  In
practice, this limit can be approached only by unusually hard sources with a
spectral index such that $F_\nu\approx$ const ($\alpha\approx 0$).

The relative intensities are also proportional to the relative escape
fractions: $Q_{\rm obs} \propto J_{\rm HeII, i}/J_{\rm H\alpha} \propto (1-
f_{\rm esc}^{\rm HeII})/(1-f_{\rm esc}^{\rm HI})$; thus, unequal escape
fractions would lead to incorrect inferred $Q$. However, as long as the escape
fraction for both HI and HeII ionizing photons is small, $f_{\rm esc} < 10 \%$,
this is a small effect.  Furthermore, while the inferred values of
$\dot{N}^{HI}_{\rm ion}, \dot{N}_{\rm ion}^{\rm HeII}$ from $J_{\rm H\alpha},
J_{\rm HeII, i}$ are subject to uncertainties about the electron temperature
and dust scattering, their ratio is much more robust. In particular, the
inferred value of Q has only a $T^{-0.15}$ dependence on the electron
temperature and is largely insensitive to interstellar reddening (Garnett et al
1991).

\section{Number of Detectable Halos}

It is interesting to ask how many sources are expected to be bright enough for
detection in HeII emission by {\it NGST}.  Based on equation~\ref{eq:SN}, the
minimum halo mass associated with a detectable source is at least $\sim
10^9~{\rm M_\odot}$.  The gas in these halos have virial temperatures $T_{\rm
vir} > 10^{4}$K, allowing efficient atomic line cooling and subsequent star or
BH formation.  We now make further simple assumptions about the efficiency of
star and BH formation to compute the expected number counts, based on the
abundance of dark halos.  The mass function of halos has recently been
determined accurately (to within $\sim20\%$) in large three--dimensional
simulations (Jenkins et al. 2000), and we adopt their fitting formula (their
eq.~9).  The halo masses relevant here are factor of a few above the ``knee''
of the mass function, where the abundance of halos is slightly larger than in
the standard Press-Schechter (1974) theory. We also adopt a flat cold dark
matter (CDM) cosmology with a cosmological constant, i.e. $\Lambda$CDM with
$(\Omega_\Lambda,\Omega_{\rm m}, \Omega_{\rm b},h,\sigma_{\rm
8h^{-1}},n)=(0.7,0.3,0.04,0.7,1,1)$.

The estimate of the number of detectable metal--free starbursts is difficult,
since the length of the epoch of completely metal--free star formation is not
known. Even trace enrichment of star forming regions to low metallicities $Z
\sim 10^{-3}$ results in stars with significantly softer spectra (Tumlinson \&
Shull 2000), with a production rate of HeII ionizing photons lower by 5 orders
of magnitude, from which HeII recombination radiation would be unobservable. It
is possible that such trace enrichment could take place very quickly, on
timescales of order the lifetime of the first generation of stars. To
circumvent this uncertainty, we simply assume that some fixed fraction $f_{*}$
of the baryonic mass of every halo with $T_{\rm vir} > 10^{4}$K undergoes a
metal--free starburst. Of course, in reality $f_{*}(M_{\rm halo},z)$ is likely
a function of both halo mass and redshift, and in particular declines with
redshift as stars pollute the IGM with metals. Our ansatz is therefore valid
only at high redshift $z > z_{\rm pollute}$ before metal pollution becomes
widespread; at lower redshifts the counts are overestimates. For detection with
NGST to be feasible $z_{\rm pollute}$ must be low, either because metal mixing
is inefficient or widespread star formation only takes place at low
redshift. We define the efficiency parameter $\epsilon=(f_{*}/0.10) (Q/0.05)$,
and in Figure~\ref{fig:stars} we show the number of detectable objects in the
three HeII lines, as a function of redshift, for two different values of
$\epsilon=1$ and $0.1$.  We require S/N=10, and adopt the field of view of
$4^\prime \times 4^\prime$ of {\it NGST}. The curves are obtained by computing
the total abundance of halos brighter than the S/N threshold, and multiplying
by the duty--cycle $t_{*}/t_H(z)$, where $t_H(z)$ is the age of the universe at
redshift $z$, and $t_*=10^{7}$ or $t_*=10^{6}$ years is the duration of the
burst. Note that since the underlying mass function is steeper than $M^{-1}$,
shorter, and therefore for a fixed $f_{*}$ brighter, bursts (upper set of three
curves) yield a larger number of detectable sources than longer bursts (middle
set of three curves).  Note further that for a low star formation efficiency of
$f_*=1\%$ (and $Q=0.05$), one source is still detectable per field out to
$z\sim13$ in the brightest line (lower solid curve).

\myputfigure{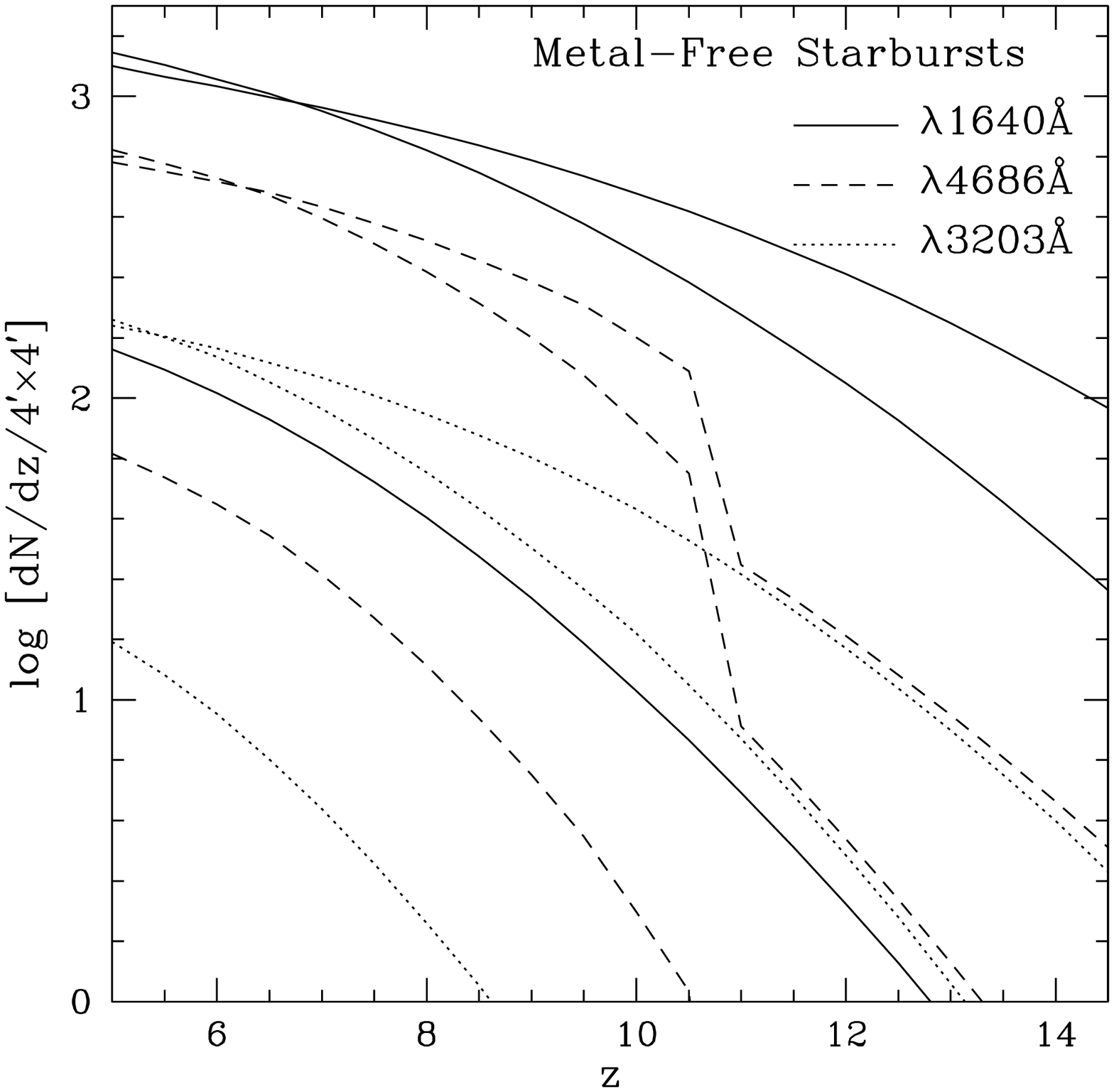}{3.2}{0.45}{-10}{-10} \figcaption{The number of sources
in a $4^{\prime} \times 4^{\prime}$ {\it NGST} field per unit redshift,
detectable as a 10$\sigma$ fluctuation in the three brightest HeII line, in a
$t=10^{5}$s exposure at spectral resolution $R=1000$. The sources are assumed
to remain unresolved. The middle set of three (solid, dashed and dotted) curves
correspond to the efficiency parameter $\epsilon=(f_{*}/0.10)(Q/0.05)=1$ and a
burst duration of $t_*=10^7$ years.  The upper and lower set of three curves
show variations from this model when $t_*=10^6$ years or $\epsilon=0.1$ is
assumed, respectively. The sharp cutoff in the counts for the $\lambda 4686\AA$
line for $z> 11$ is due to the increased detector noise at $\lambda > 5.5 \,
\mu$m.
\label{fig:stars}}
\vspace{\baselineskip}

To estimate the number of detectable mini--quasars, we assume that a fraction
$3\times 10^{-4}(\Omega_{\rm b}/\Omega_{\rm m})$ of the halo mass collects into
a central BH, which shines for $t_Q=10^7$ years.  Alternatively, we assume that
the BH's are 10 times more massive than this, but shine only for $10^6$
years. Both of these relations produce the same amount of total quasar light,
and the latter relation is similar to the Haiman \& Loeb (1998) model, which
was explicitly calibrated to fit the Pei (1995) luminosity function at
redshifts below $z \lsim 5$ (see Haehnelt et al. 1998 and Haiman \& Hui 2000
for a discussion on the quasar lifetime, and the calibration of the ratio of BH
to halo mass). Note that while the low-redshift empirical relation $M_{\rm bh}
\propto L_{\rm bulge}$ (Magorrian et al 1998) has large scatter, the relation
$M_{\rm bh} \propto \sigma^{\alpha}$ has recently been shown to be much
tighter, with scatter largely accounted for by observational errors (Gebhardt
et al 2000, Ferrarese et al 2000). This new result is in line with our
assumptions, since at a given formation redshift the correlation $M_{\rm halo}
\propto \sigma^{\alpha}$ is much tighter than the correlation $M_{\rm halo}
\propto {\rm L_{bulge}}$ (the latter being subject to larger scatter in gas
cooling and star formation efficiency). Nevertheless, we must regard the
$M_{\rm bh} \propto M_{\rm halo}$ scaling as an uncertain ansatz motivated by
lower--redshift observations. In Figure~\ref{fig:qsos} we show the expected
number of detectable sources in these two models in the three strongest HeII
lines.  Similar to the starburst case, shorter and brighter mini--quasars would
be more numerous, especially at the highest redshifts, where the underlying
halo mass function is steep (cf. the upper vs. the middle set of three curves).
The lower set of three curves demonstrate that for a relatively soft spectrum
$\alpha=1.8$, a few mini--quasars would still be visible out to $z\sim 14$.

\myputfigure{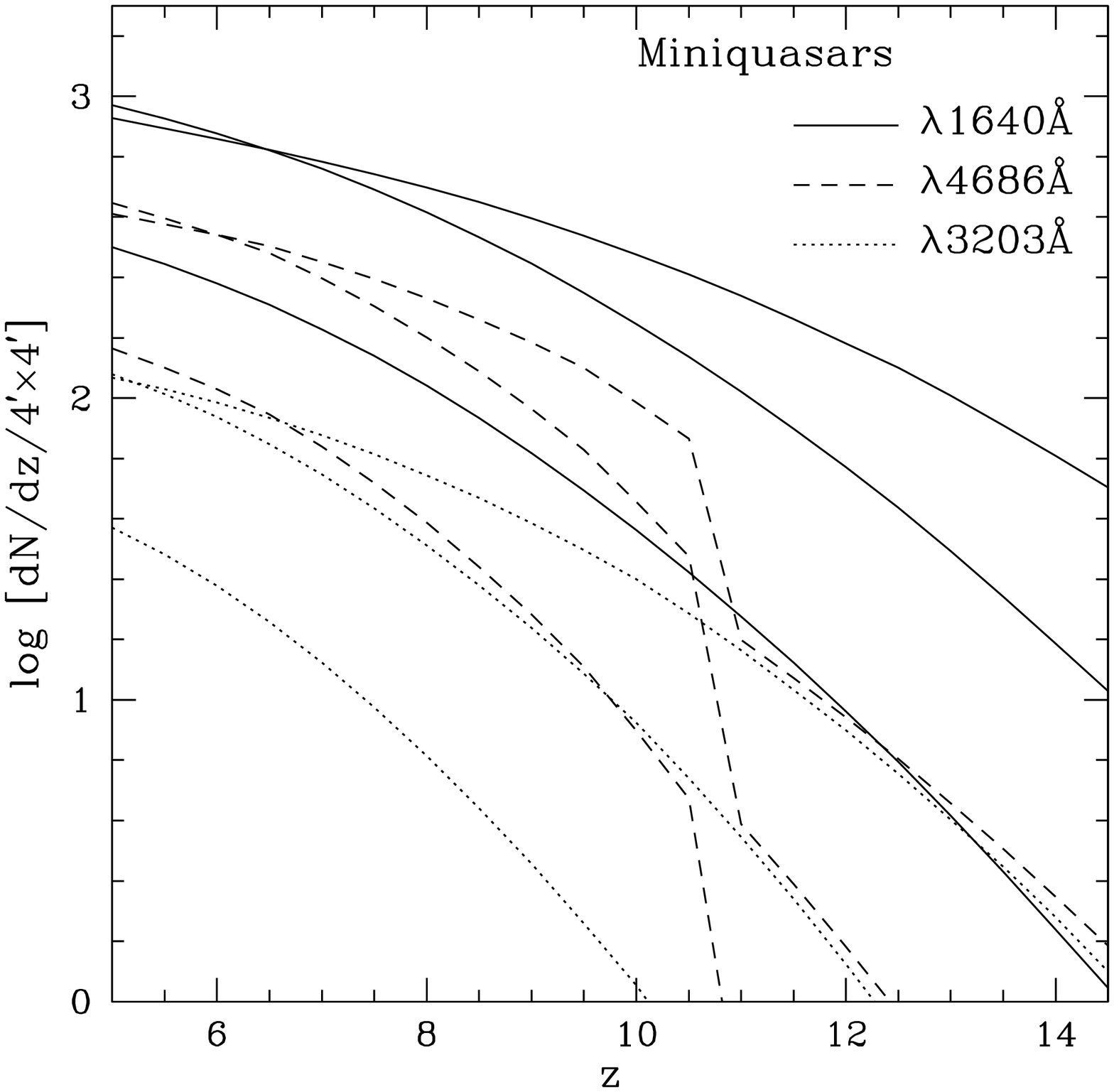}{3.2}{0.45}{-10}{-10} \figcaption{Same as
Fig.~\ref{fig:stars}, but for mini--quasars. The middle set of three curves
describe a model with BH to halo mass ratio of $3\times 10^{-4}(\Omega_{\rm
b}/\Omega_{\rm m})$, a quasar lifetime of $t_Q=10^7$ years, and spectral index
$\alpha=1$. The upper and lower set of three curves show variations from this
model when $t_Q=10^6$ years (with $M_{\rm BH}/M_{\rm halo}=3\times
10^{-3}[\Omega_{\rm b}/\Omega_{\rm m}]$) or $\alpha=1.8$ is assumed,
respectively.
\label{fig:qsos}}

\section{Discussion}

Our simple estimates suggest that hard ionizing radiation from either
metal--free stellar populations or from mini--quasars would produce HeII
emission lines that are detectable with {\it NGST}.  Our results, shown in
Figures~\ref{fig:stars} and~\ref{fig:qsos} reveal that there could be $\sim 10$
suitably bright sources from redshift $z\gsim 10$ in an {\it NGST} field in a
$10^5$ sec exposure. The distinction between metal--free starbursts and mini--quasars
is best performed with observations with greater signal-to-noise than HeII
recombination lines--e.g. on the basis of broad-band colors as has been
successfully used to select high-redshift quasars (Fan 1999), H$\alpha$ line
widths (with mini--quasars displaying much broader line widths $\sigma \sim
1000 \, {\rm km s^{-1}}$ due to line broadening by the accretion disc), and the
fact that the brightest mini-quasars can be seen in X-ray emission with CXO
(see discussion below).

HeII recombination radiation could also arise from other sources of hard
photons, such as supernova remnants.  Indeed, HeII recombination emission has
been detected in several nearby extragalactic HII regions, and have been used
to infer values of $Q$ of order $Q \approx 3 \%$ (e.g., Garnett et al
1991). The interpretation of these results is still uncertain, with suggested
emission mechanisms for HeII ionizing radiation ranging from Wolf-Rayet stars
to X-ray binaries and shocks. Suffice it to say that if such hard sources are
also present at high redshift, their HeII recombination will also be detectable
by {\it NGST}, and their presence has important implications for the
reionization process. For processes which scale with the star formation rate
such as Wolf-Rayet stars, XRB and supernova shocks, their detectability can be
estimated simply by inserting the appropriate value of $Q$ in equation
(\ref{eq:Jstar}); the empirically derived value of $Q \approx 3 \%$ yields
fluxes about a factor of 2 below that assumed for metal-free stars. Thus, HeII
recombination lines might be visible from star-forming regions even after
widespread metal pollution.

Candidate objects for HeII emission lines can be imaged with {\it NGST} in
broad bands at rest-frame energies below 13.6eV. However, strong absorption
within the source, and by the neutral IGM, will render the ionizing continuum
inaccessible to direct observation.  High--redshift quasars are also much more
difficult to detect in X--rays than in the HeII lines.  To detect a redshift
$z=10$ source with the {\it Chandra} satellite in $5\times 10^5$ seconds at
S/N=5, the BH mass needs to be at least about $10^8~{\rm M_\odot}$ (Haiman \&
Loeb 1999). By contrast, {\it NGST} could detect the HeII line signal from a BH
as small as $\sim 10^5~{\rm M_\odot}$, assuming the same redshift, S/N, and
duration of observation (cf. eq.~\ref{eq:SN}). Unless nearly all ionizing
radiation ($f_{\rm esc}\gsim 99.9\%$) leaks out from high--redshift quasars,
the HeII detection with {\it NGST} is therefore easier than the X-ray
observation with {\it Chandra}.

It is worth noting that the measured value of Q does not uniquely constrain the
importance of hard photons for the reionization of the universe.  Photons with
energies above $E \gsim 270 (N_{\rm HI}/10^{21}{\rm cm^{-2}})^{1/3}$eV can
escape unimpeded from a halo with a neutral hydrogen column density of $N_{\rm
HI}$, and would not produce HeII recombinations (and would therefore not affect
the measurement of Q).  Nevertheless, these hard photons can contribute to the
reionization of the universe by multiple secondary ionizations far away from
the source, since a fully neutral IGM is optically thin only for $E \gsim 1.5
\redshiftt^{1/2} \langle x_{\rm HI}\rangle^{1/3}$keV. Such photons are more
important for HI ionizations rather than HeII ionizations, since each photon
contributes at most one HeII ionization but multiple HI ionizations (secondary
ionizations have little effect on HeII ionization, Shull \& van Steenburg
1985).  Thus, hard sources of radiation, such as non-thermal emission from
supernova remnants, can produce low levels of HeII recombination line flux and
yet potentially play an important role in the reionization of the universe (Oh
2000a). For these sources, the best observational handle on hard photon
production is radio synchrotron emission, which arises from the same
relativistic electron population which produces X-ray emissions.

\section{Conclusions}

The unparalleled sensitivity of {\it NGST} will make it possible to image the
first luminous sources in near and mid IR, corresponding to rest frame UV
emission longward of the Lyman limit (e.g. Haiman \& Loeb 1997; 1998). However,
the most important part of the spectrum for models of reionization, the
ionizing continuum, will likely be inaccessible to direct observation. Provided
that the escape fraction of ionizing photons is indeed small, the recombination
line fluxes can serve as an important proxy for this part of the spectrum.
This technique has already been successfully applied to extragalactic HII
regions (Garnett 1991). The Balmer $\alpha$ flux is relatively easily
detectable from high--redshift sources (Oh 1999), and probes their intrinsic
spectrum in the range $13.6-54.4$eV. In this paper, we have shown that the HeII
recombination lines are also detectable for redshifts exceeding $z\sim 10$, and
provide a measure of the source luminosity above $54.4$eV.  The most promising
line is HeII$(3 \rightarrow 2)$ ($1640 {\rm \AA}$), with the HeII$(4 \rightarrow 3)$ ($4686 {\rm \AA}$) and HeII$( 5
\rightarrow 3)$ ($3203 {\rm \AA}$) lines fainter by a factor of $\sim
2$ and $\sim 8$ respectively.  Combining the measured H$\alpha$ and HeII fluxes can serve as an
important probe of the hardness of the spectrum, largely independent of the
electron temperature and interstellar reddening.  Such a measurement will shed
light on the nature and emission mechanism of these first luminous sources,
with important astrophysical consequences for the reheating and reionization of
the IGM.

\acknowledgments

SPO acknowledges support by NASA through ATP grant NAG5-7154. ZH was supported
by NASA through the Hubble Fellowship grant HF-01119.01-99A, awarded by the
Space Telescope Science Institute, which is operated by the Association of
Universities for Research in Astronomy, Inc., for NASA under contract NAS
5-26555.  We thank the anonymous referee for helpful comments.

\end{document}